\documentclass{article}

\usepackage[preprint]{neurips_2024}

\usepackage[utf8]{inputenc} 
\usepackage[T1]{fontenc}    
\usepackage{hyperref}       
\usepackage{url}            
\usepackage{booktabs}       
\usepackage{amsfonts}       
\usepackage{nicefrac}       
\usepackage{microtype}      

\usepackage[utf8]{inputenc} 
\usepackage[T1]{fontenc}    
\usepackage{hyperref}       
\usepackage{url}            
\usepackage{booktabs}       
\usepackage{amsfonts}       
\usepackage{nicefrac}       
\usepackage{microtype}      
\usepackage[dvipsnames,svgnames,table]{xcolor}         
\usepackage{amsmath}
\usepackage{mathtools}
\usepackage{cleveref}
\usepackage{amsthm}
\usepackage{amssymb}
\usepackage{algorithm}
\usepackage[noend]{algpseudocode}
\usepackage{comment}
\usepackage{mathtools}
\usepackage{wrapfig}
\usepackage{bm}
\usepackage[normalem]{ulem}

\usepackage[shortlabels]{enumitem}
\setenumerate[1]{leftmargin=2em, itemsep=0pt, partopsep=0pt, parsep=\parskip, topsep=0pt}
\setenumerate[2]{leftmargin=2em, itemsep=0pt, partopsep=0pt, parsep=\parskip, topsep=0pt}
\setitemize[1]{leftmargin=2em, itemsep=0pt, partopsep=0pt, parsep=\parskip, topsep=0pt}
\setitemize[2]{leftmargin=2em, itemsep=0pt, partopsep=0pt, parsep=\parskip, topsep=0pt}

\usepackage{tikz}
\usetikzlibrary{arrows.meta, shapes}

\newtheorem{theorem}{Theorem}[section]
\newtheorem{lemma}[theorem]{Lemma}

\newtheorem{proposition}{Proposition}[section]
\newtheorem{corollary}[theorem]{Corollary}

\usepackage{etoolbox}
\pretocmd{\lemma}{\crefalias{theorem}{lemma}}{}{}
\pretocmd{\corollary}{\crefalias{theorem}{corollary}}{}{}
\pretocmd{\proposition}{\crefalias{theorem}{proposition}}{}{}
\Crefname{assumption}{Assumption}{Assumptions}

\newcommand{\norm}[1]{\left\lVert#1\right\rVert}

\newcommand{\MPC}{\mathsf{MPC}}
\newcommand{\OPT}{\mathsf{OPT}}
\newcommand{\ALG}{\mathsf{ALG}}
\newcommand{\cost}{\mathrm{cost}}

\allowdisplaybreaks

\title{Empirical Study of Dynamic Regret in Online Model Predictive Control for Linear Time-Varying Systems}

\author{%
  Nhat M. Nguyen \\
  Department of Computer Science\\
  University of Alberta\\
  116 St \& 85 Ave, Edmonton, AB T6G 2R3 \\
  \texttt{nmnguyen@ualberta.ca} \\
}

\begin{document}

\maketitle

\begin{abstract}
Model Predictive Control (MPC) is a widely used technique for managing time-varying systems, supported by extensive theoretical analysis. While theoretical studies employing dynamic regret frameworks have established robust performance guarantees, their empirical validation remains sparse. This paper investigates the practical applicability of MPC by empirically evaluating the assumptions and theoretical results proposed by \cite{lin2022bounded}. Specifically, we analyze the performance of online MPC under varying prediction errors and prediction horizons in Linear Time-Varying (LTV) systems. Our study examines the relationship between dynamic regret, prediction errors, and prediction horizons, providing insights into the trade-offs involved. By bridging theory and practice, this work advances the understanding and application of MPC in real-world scenarios.
\end{abstract}

\section{Introduction}
Model Predictive Control (MPC) is a well-established framework for addressing optimal control problems in time-varying systems. Its ability to handle dynamic environments has made it a widely adopted tool across applications such as robotics, autonomous systems, and energy optimization. Despite its theoretical appeal, bridging the gap between theoretical guarantees and practical implementation remains a significant challenge, particularly in the context of dynamic regret analysis for online MPC.

Dynamic regret - a performance metric that quantifies the difference between an online controller and its offline optimal counterpart - has become central to evaluating the efficiency of online MPC. Recent advancements, such as the framework proposed by \cite{lin2022bounded}, establish dynamic regret bounds using perturbation analysis, accounting for the influence of prediction errors and prediction horizons on control performance. While these results are promising, they rely on theoretical assumptions that have yet to be comprehensively validated, particularly for Linear Time-Varying (LTV) systems.

This paper addresses this gap by conducting a systematic empirical study of the dynamic regret bounds for online MPC in LTV systems. By simulating realistic control scenarios with varying prediction errors and prediction horizons, we assess the practical validity of the theoretical assumptions and insights from \cite{lin2022bounded}. Furthermore, we examine the interplay between prediction errors, prediction horizons, and resulting performance, offering a deeper understanding of the trade-offs inherent in online MPC.

The findings of this study contribute to bridging the gap between theory and practice in online MPC. By empirically validating key theoretical results and identifying their limitations, this work provides actionable insights for refining predictive control strategies and highlights future research directions, such as adaptive horizon selection for resource-constrained systems.

\section{Dynamic Regret for Online Model Predictive Control}
First, we define the online control problem and dynamic regret for online MPC, closely following the formulation in \cite{lin2022bounded}. Consider a discrete, finite-horizon, time-varying optimal control problem formulated as:
\begin{align}\label{equ:online_control_problem}
    \min_{x_{0:T}, u_{0:T-1}} &\sum_{t = 0}^{T-1} f_t(x_t, u_t; \xi_t^*) + F_T(x_T; \xi_T^*) \nonumber\\*
    \text{ s.t. }&x_{t+1} = g_{t}(x_{t}, u_{t}; \xi_{t}^*), &\forall 0 \leq t < T,\nonumber\\*
    &s_t(x_t, u_t; \xi^*_t) \leq 0, &\forall 0 \leq t < T,\\*
    &x_0 = x(0).\nonumber
\end{align}
Here: $x_t \in \mathbb{R}^n$ is the state, $u_t \in \mathbb{R}^m$ is the control input. $f_t$ is a time-varying cost function, $g_t$ is a time-varying dynamic function, and $s_t$ is a time-varying constraint function. $F_T$ represents the terminal cost function. The collection of these functions is referred to as the \textit{problem data} of the control problem, which we assume is generated by a ground truth parameter $\xi_t^*$ (unknown in practice).

An offline controller $\OPT$ that receives full knowledge of the ground truth parameter $\xi_t^*$ can solve the control problem optimally, generating the optimal trajectory $\OPT$ with the lowest cost. However, in practice, we do not have access to these ground truth parameters. At each time step $t$, we only have access to noisy predictions $\xi_{t:t+k \mid t}$ of the ground truth parameters for a fixed \textbf{prediction horizon} of $k$ future time steps. Additionally, we define the \textbf{prediction errors} at time $t$ as $\rho_{t, \tau} \coloneqq \norm{\xi_{t + \tau \mid t} - \xi_{t+\tau}^*}$ for $\tau=0 \dots k$. A controller $\ALG$ that attempts to solve the optimal control problem using these noisy predictions is called an online controller. Since the online controller lacks full knowledge of the ground truth, it incurs additional costs compared to the optimal controller $\OPT$. The objective is to design the online controller to compete against the offline controller $\OPT$. In this work, we use the performance metric \textit{dynamic regret}. Let $\cost(\OPT)$ and $\cost(\ALG)$ denote the total costs incurred by $\OPT$ and $\ALG$, respectively. The \textit{dynamic regret} for a given instance of $x(0), \xi_{0:T}^*$ is defined as the worst-case additional cost incurred by $\ALG$ compared to $\OPT$:  
$\sup_{x(0), \xi_{0:T}^*} (\cost(\ALG) - \cost(\OPT))$.

A popular framework for solving such online optimal control problems is Model Predictive Control (MPC). MPC solves the optimization problem in \ref{equ:online_control_problem} using the predicted parameters under a fixed prediction horizon, committing the first control input, receiving feedback, and iterating this process until the terminal time step $T$. This is called online MPC. When MPC receives full knowledge of the ground truth parameters, the algorithm becomes optimal, referred to as offline MPC.

The work by \cite{lin2022bounded} proposes a three-step pipeline to derive dynamic regret bounds for online MPC. Below is a summary of the three steps (for complete details, refer to the original paper):

\textbf{Step 1: Obtain perturbation bounds in the following forms:}
\begin{enumerate}[nosep,leftmargin=.2in,label=(\alph*)]
    \item \textit{Perturbation of the parameters $\xi_{t_1:t_2}$ given a fixed initial state $z$}:
    \begin{equation}\label{equ:perturbation-bound-fix-initial}
        \norm{\psi_{t_1}^{t_2}\left(z, \xi_{t_1:t_2}; F\right)_{v_{t_1}} - \psi_{t_1}^{t_2}\left(z, \xi_{t_1:t_2}'; F\right)_{v_{t_1}}} \leq \left(\sum_{t=t_1}^{t_2} q_1(t - t_1) \delta_t\right) \norm{z} + \sum_{t=t_1}^{t_2} q_2(t - t_1) \delta_t,
    \end{equation}
    where $\delta_t \coloneqq \norm{\xi_t - \xi_t'}$ for $t \in [t_1, t_2]$, and scalar functions $q_1$ and $q_2$ satisfy
    $\lim_{t\to\infty}q_i(t) = 0$, $\sum_{t=0}^\infty q_i(t) \leq C_i$ for constants $C_i \geq 1$, $i = 1, 2$.
    \item \textit{Perturbation of the initial state $z$ given fixed parameters $\xi_{t_1:t_2}$}:
    \begin{equation}\label{equ:perturbation-bound-fix-parameters}
        \norm{\psi_{t_1}^{t_2}\left(z, \xi_{t_1:t_2}; F\right)_{y_t/v_t} - \psi_{t_1}^{t_2}\left(z', \xi_{t_1:t_2}; F\right)_{y_t/v_t}} \leq q_3(t - t_1) \norm{z - z'}, \text{ for } t \in [t_1, t_2],
    \end{equation}
    where the scalar function $q_3$ satisfies $\sum_{t=0}^\infty q_3(t) \leq C_3$ for some constant $C_3 \geq 1$.
\end{enumerate}

The first perturbation bound controls how much the control input chosen by the online MPC controller can deviate when the parameters are perturbed. The second perturbation bound controls how much the state and control input trajectories chosen by the online MPC can deviate when the initial state at each time step $t$ is changed. From the original paper, we observe that the right-hand sides of these perturbation bounds approach zero as the prediction errors decrease and the prediction horizon $k$ increases.

\textbf{Step 2: Bound the per-step error $\bm{e_t}$.} The per-step error is defined as the difference between the distance between the online controller $\ALG$ and the offline optimal action: $e_t \coloneqq \norm{u_t - \psi_t^T(x_t, \xi_{t:T}^*; F_T)_{v_t}}, \text{ where }u_t = \ALG(x_t, \xi_{t:t+k\mid t})$. Under some assumptions and using the perturbation bounds in step 1, we can bound the per-step error by:

\begin{equation}\label{lemma:pipeline-step2:conclusion}
    e_t \leq \sum_{\tau = 0}^{k}\left(\left(\frac{R}{C_3} + D_{x^*}\right) \cdot q_1(\tau) + q_2(\tau)\right)\rho_{t, \tau} + 2R\left(\left(\frac{R}{C_3} + D_{x^*}\right) \cdot q_1(k) + q_2(k)\right)
\end{equation}

\textbf{Step 3: Bound the dynamic regret by $\bm{\sum_{t=0}^{T-1} e_t^2}$}: Under additional assumptions such that $e_t \leq \frac{R}{C_3^2 L_g}$ for all $t < T$, the dynamic regret of MPC is upper-bounded by: $\cost(\ALG) - \cost(\OPT) = O\left(\sqrt{\cost(\OPT)\cdot \sum_{t=0}^{T-1} e_t^2} + \sum_{t=0}^{T-1} e_t^2\right)$. With a long enough prediction horizon and low enough prediction errors, the dynamic regret can be sublinear.

\section{Related Work}

Numerous studies have focused on deriving performance guarantees for Model Predictive Control (MPC). \cite{yu2020power} established constant dynamic regret for online MPC under perfect predictions and logarithmic prediction horizons, while \cite{zhang2021regret} demonstrated exponential decay of dynamic regret with increasing prediction horizons, though their empirical validation was limited. \cite{lin2021perturbation} introduced perturbation bounds for Linear Time-Varying (LTV) systems, leveraging them to prove competitive ratios and dynamic regret bounds. Similarly, works by \cite{shin2020decentralized}, \cite{shin2021controllability}, \cite{xu2019exponentially}, and \cite{na2022superconvergence} advanced perturbation analysis methods for online control, providing valuable theoretical insights.

Building on these, \cite{lin2022bounded} proposed a systematic framework that generalizes prior results, offering flexible performance guarantees for online MPC whenever perturbation bounds can be proven. The authors also used their framework to derive new theoretical results. However, like most prior work, this study lacks empirical validation, highlighting a general weakness in the literature. Thus, there is a critical need for a thorough empirical evaluation of performance guarantees to assess their practical applicability in real-world scenarios.

\section{Experimental Setup}

\subsection{LTV System Formulation for Experiments}
\label{section:LQR_formulation}
\begin{figure}[h!]
    \centering
    \includegraphics[width=0.8\textwidth]{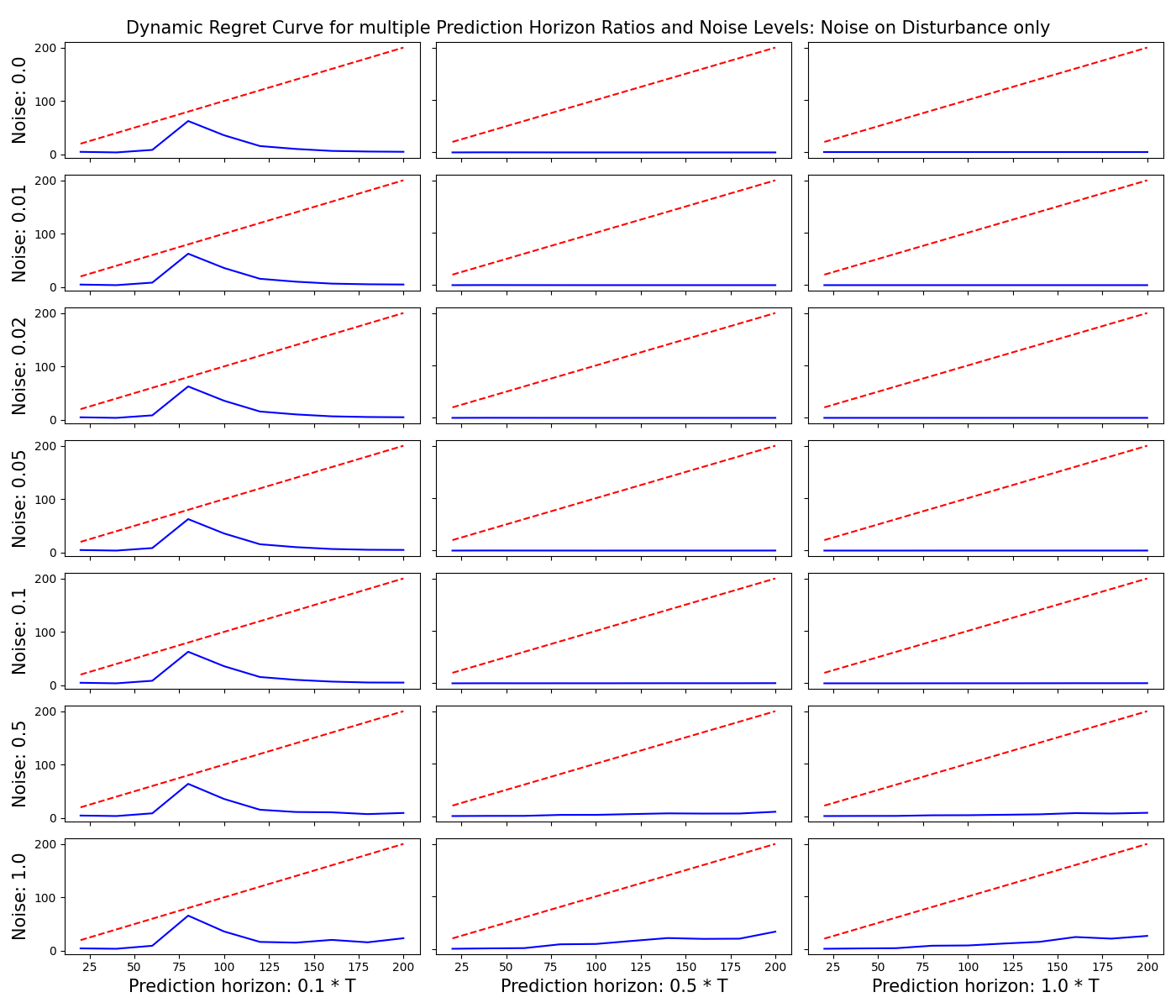} 
    \caption{Empirical dynamic regret when prediction errors only affect the disturbance. \textbf{For each subplot, the x-axis represents episode length $T$, ranging from 20 to 200.}}  \label{fig:regret_disturbance} 
\end{figure}

In this work, we provide empirical evidence supporting the results of \cite{lin2022bounded} for Linear Time-Varying (LTV) systems. Due to limitations with the CVXPY solver, we were unable to perform experiments on nonlinear systems. For all experiments, we consider the Linear Quadratic Regulator (LQR) problem with disturbances in a discrete, finite horizon, time-varying system setting, as follows:
\begin{align}\label{equ:online_control_problem:unconstrained-LTV-dynamics}
    \min_{x_{0:T}, u_{0:T-1}} &\sum_{t = 0}^{T-1} \left( (x_t - \bar{x}_t(\xi_t^*))^\top Q_t(\xi_t^*) (x_t - \bar{x}_t(\xi_t^*)) + u_t^{\top} R_t(\xi_t^*) u_t \right) + F_T(x_T; \xi_t^*)\nonumber\\*
    \text{ s.t. }&x_{t+1} = A_t(\xi_{t}^*)\cdot x_{t} + B_t(\xi_{t}^*)\cdot u_{t} + w_t(\xi_{t}^*), \hspace{6em}\forall 0 \leq t < T,\\*
    &x_0 = x(0), \nonumber
\end{align}
where the terminal cost is given by $F_T(x_T; \xi_T^*) \coloneqq (x_T - \bar{x}_T(\xi_T^*))^\top P_T(\xi_T^*) (x_T - \bar{x}(\xi_T^*))$.

In our experiments, the cost matrices are defined as follows:
$Q_t(\xi^*_t) =
\begin{bmatrix}
1 + \exp(-\tilde{t}) & 0 \\
0 & 1 + 0.05 \cdot \tilde{t}
\end{bmatrix}$,
$R_t(\xi^*_t) =
\begin{bmatrix}
1 & 0 \\
0 & 1 + \exp(-\tilde{t})
\end{bmatrix}$, 
$\bar{x}_t(\xi^*_t) = 
\begin{bmatrix}
\sin(\tilde{t}) \\
\cos(\tilde{t})
\end{bmatrix}
$, 
$P_T(\xi^*_T) =
\begin{bmatrix}
1 + \exp(-\tilde{T}) & 0 \\
0 & 1 + 0.05 \cdot \tilde{T}
\end{bmatrix}
$.

The transition matrices are:
$A_t(\xi^*_t) =
\begin{bmatrix}
\cos(\tilde{t}) & \sin(\tilde{t}) \\
-\sin(\tilde{t}) & \cos(\tilde{t})
\end{bmatrix}
$,
$B_t(\xi^*_t) =
\begin{bmatrix}
1 & 0 \\
0 & 0.1 + \exp(-\tilde{t})
\end{bmatrix}
$.
The disturbance is modeled as noise sampled from Gaussian distributions at each time step:
$w_t(\xi^*_t) = 
\begin{bmatrix}
w^1_t \sim \mathcal{N}(0, 0.2) \\
w^2_t \sim \mathcal{N}(0, 0.2)
\end{bmatrix}
$. 
Here, $\tilde{t} = t \Delta t$ with $\Delta t = 0.1$ as the system time step, and $\tilde{T} = T \Delta t$. $T$ is the total number of time steps (or episode length). This relationship reflects the translation between system time steps and discrete time steps of the optimization problem, where 0.1 seconds in the system corresponds to one discrete time step. This LQR system formulation is widely used in the literature and has practical applications.

\textbf{Addressing the required assumptions of LQR formulation:} This LQR formulation satisfies many of the assumptions in the work of \cite{lin2022bounded}, such as Lipschitzness, stability, boundedness, $\mu$-strongly convexity and $l$-smoothness of the cost functions. We also found that, empirically, recursive feasibility is also satisfied in all of our experiments. We haven't found any case where the LQR system failed to be solved in all of our runs. Controllability can also be proved by reasoning with intuition: For any target state that we want to achieve, consider the matrix 
$B_t(\xi^*_t)$, we can see that the contribution of the control action to the changing of the next state is at least $1.0$ and $0.1$ for both states and there is no constraint on the control input in this LQR formulation. Therefore, we can increase or decrease the chosen control input arbitrarily to achieve any desired target state. Thus, we posit that this system satisfies all the assumptions for LTV systems in the work of \cite{lin2022bounded} and can be used for empirical validation of the results in the paper.

In this paper, the dynamic regret bound is defined as the upper bound of total cost difference over all conditions for a starting state $x(0)$ and the sets of parameters $\xi_{0:T}^*$. This is intractable. Instead of attempting to validate the dynamic regret bound directly, we study empirical dynamic regret $[\cost(\MPC) - \cost(\OPT)]_{x(0),\xi_{0:T}^*}$ for a single starting state $x(0)$ and a single set of parameters $\xi_{0:T}^*$ under various experiment settings. One key point of the results in \cite{lin2022bounded} is that there are trade-off relationships between prediction errors $e_t$ and prediction horizon $k$ for ensuring sublinear dynamic regret bound. Therefore, we designed our experiments such that we can study the relationship between the dynamic regret bound against various levels of prediction errors and prediction horizons. We implemented the system formulated in section~\ref{section:LQR_formulation} using CVXPY (\cite{diamond2016cvxpy}, \cite{agrawal2018rewriting}) under fixed random seeds (to fix the random disturbances). Then, we ran a clairvoyant offline MPC controller to solve the system and recorded all the states, control inputs, all problem data, and all intermediate steps of the optimization for a certain episode length $T$. Then, we implemented an online MPC controller with varying levels of prediction errors added under short to long prediction horizons. Again, we recorded everything just like in the offline run. Then, we performed analyses using the recorded data and plotted all the relevant information. Additionally, because of the differences in results and assumptions for LTV systems in the work of \cite{lin2022bounded}, we split our experiment into two settings similar to those in the mentioned paper: \textbf{(1)} When there are only predictions errors on the disturbance $w_t(\xi_t^*)$ and \textbf{(2)} When there are prediction errors on all the problem data of the online optimization.

\subsection{Methodology for Controlling Prediction Errors}

A key aspect of our empirical validation of the dynamic regret bound is controlling various levels of prediction errors. In \cite{lin2022bounded}, prediction errors are quantified with respect to the parameters $\xi_t^*$, which define the problem data of the optimization. In our experiments, we model prediction errors directly on the problem data of the optimization. Specifically, we treat the parameters $\xi_t^*$ as the collection of all problem data at each time step $t$. To simulate prediction errors, we add uniform noise sampled from $U(-\epsilon, \epsilon)$ to the relevant problem data for all time steps.
For example, in experiment setting \textbf{(1)}, where prediction errors only affect the disturbance, the disturbance for the online MPC controller is modified as follows: $\omega_t(\xi_t) = \omega_t(\xi_t^*) + U(-\epsilon, \epsilon)$,
where $U(-\epsilon, \epsilon)$ represents the uniform distribution, and $\epsilon$ is the \textbf{noise strength}. In experiment setting \textbf{(2)}, uniform noise is added to all problem data, including $Q_t(\xi_t^*)$, $R_t(\xi_t^*)$, $\bar{x}_t(\xi_t^*)$, $P_T(\xi_T^*)$, $A_t(\xi_t^*)$, $B_t(\xi_t^*)$, and $w_t(\xi_t^*)$.
The noise strength parameter $\epsilon$ serves as a control for the magnitude of the prediction errors. In our experiments, we vary $\epsilon$ across the values $[0.0, 0.01, 0.02, 0.05, 0.1, 0.5, 1.0]$, where $\epsilon = 0.0$ indicates no prediction errors, and $\epsilon = 1.0$ corresponds to the maximum level of prediction errors.

\section{Empirical Results}
\subsection{Empirical Dynamic Regret of the Proposed LQR System}

\begin{figure}[h!]
    \centering
    \includegraphics[width=0.8\textwidth]{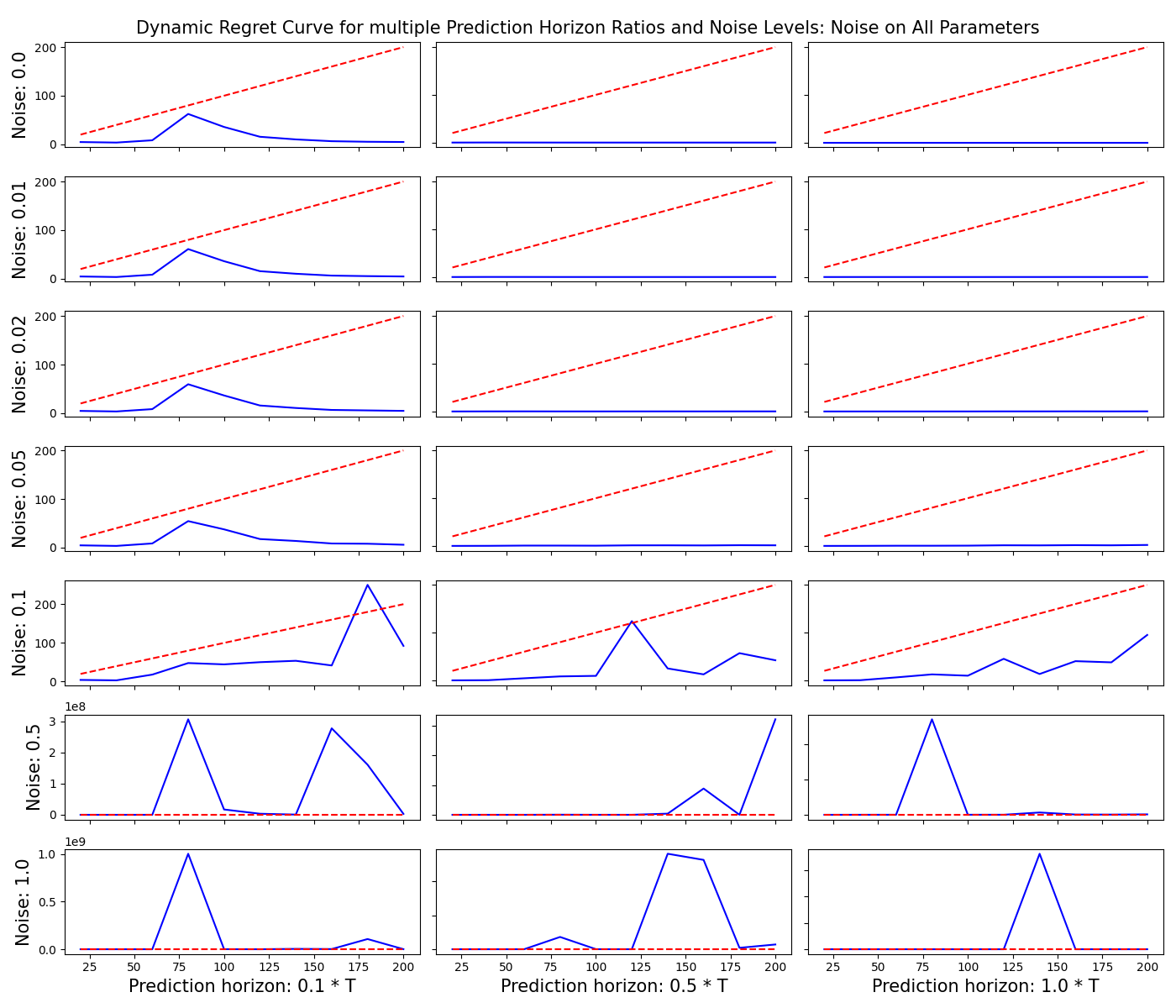} 
    \caption{Empirical dynamic regret when prediction errors are present in all problem data of the optimization. \textbf{For each subplot, the x-axis represents episode length $T$, ranging from 20 to 200.}}
    \label{fig:regret_all} 
\end{figure}

First, we performed experiments to validate the empirical dynamic regret of the LQR system under various prediction horizons and levels of prediction errors. According to Section 4 in \cite{lin2022bounded}, the following hypotheses were made: (1) When prediction errors are only present in the disturbance, sublinear dynamic regret can be achieved for any level of prediction error, provided the prediction horizon is sufficiently large; (2) When prediction errors are present in all problem data, there is a trade-off between prediction accuracy and prediction horizon for achieving sublinear dynamic regret - larger prediction errors necessitate longer prediction horizons.

To validate these theoretical hypotheses, we ran both the offline and online MPC controllers for increasing episode lengths, $T$, ranging from 20 to 200 total time steps. We varied the level of noise strength of the uniform distribution, as described in the previous section. Additionally, we considered three values of prediction horizon: $0.1 T$ (short prediction horizon - 10\% of episode length), $0.5 T$ (long prediction horizon - 50\% of episode length), and $1.0 T$ (full prediction horizon - 100\% of episode length). This formulation ensures that the results are meaningful and that the prediction horizon satisfies $k = \Omega(T)$, as required to achieve sublinear dynamic regret, according to Theorem 3.3 in \cite{lin2022bounded}.

Figure~\ref{fig:regret_disturbance} shows the empirical dynamic regret when prediction errors are present only in the disturbance. We observe that sublinear dynamic regret is achieved across all levels of prediction errors and all prediction horizons for episode lengths $T$ ranging from 20 to 200. For maximum noise strength, dynamic regret increases as the episode length $T$ increases. However, it remains sublinear even at $T = 200$. For the short prediction horizon of $0.1 T$, a spike in dynamic regret is observed between episode lengths $T$ of 60 and 120. We hypothesize that this spike occurs because the terminal cost function in our LQR formulation is myopic: it attempts to match the next reference state without considering the longer-term trajectory. This can cause the controller to optimize control inputs that are effective in the short term but detrimental in the long term, particularly if the reference trajectory changes drastically after the terminal reference state within the prediction horizon. This effect disappears as the episode length $T$ exceeds 120, allowing the online MPC controller to account for the changing reference trajectory beyond the horizon.
Figure~\ref{fig:regret_all} presents the empirical dynamic regret when prediction errors are present across all problem data. Unlike the previous case, sublinear dynamic regret is only achieved for combinations where the prediction errors are sufficiently low, and the prediction horizon is long enough.

These empirical results align with the theoretical expectations of dynamic regret outlined earlier in this section. Therefore, we can conclude that our experiments provide empirical evidence supporting the dynamic regret bound presented in \cite{lin2022bounded}.

\subsection{Exponential Decay of Per-Step Error with Prediction Horizon}

\begin{figure}[h!]
    \centering
    \begin{minipage}{0.48\textwidth}
        \centering
        \includegraphics[width=\textwidth]{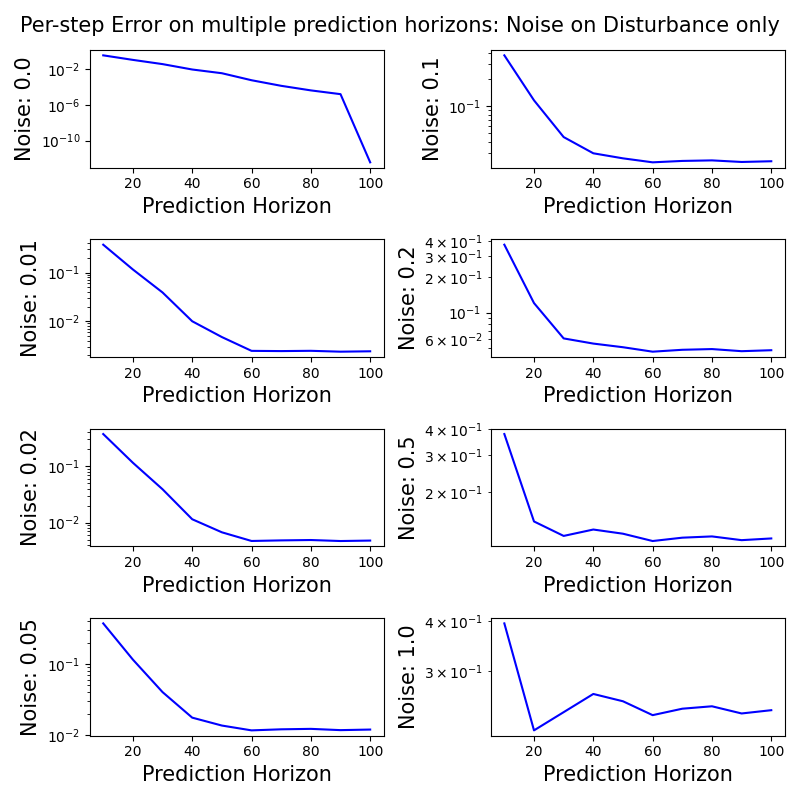}
        \label{fig:per-step_errors1}
    \end{minipage}
    \hfill
    \begin{minipage}{0.48\textwidth}
        \centering
        \includegraphics[width=\textwidth]{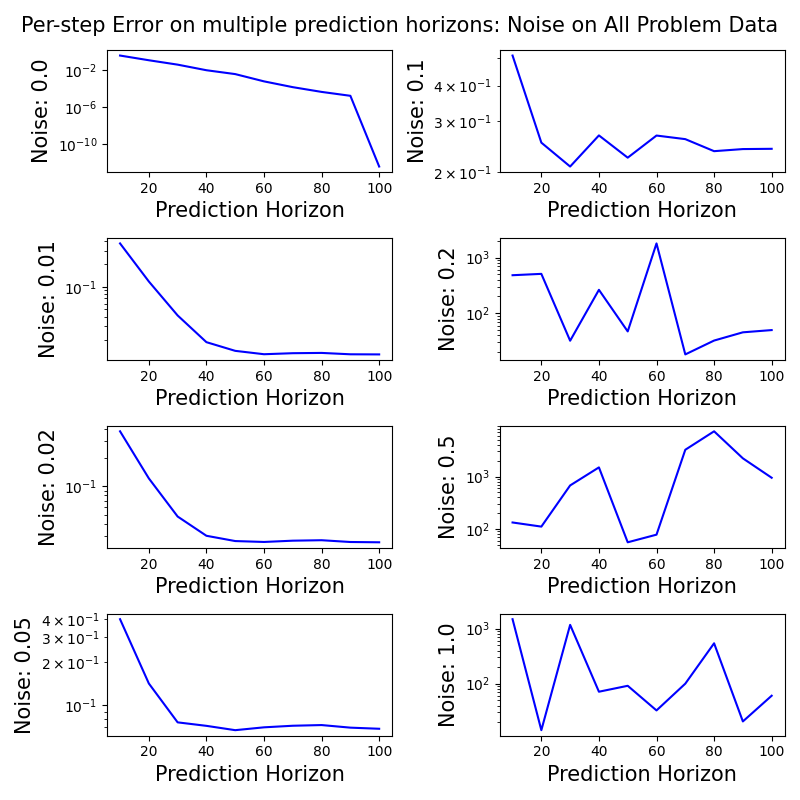}
        \label{fig:per-step_errors2}
    \end{minipage}
    \caption{Per-step error (log scale) for cases with prediction errors on disturbance only (left) and on all problem data (right). For noise strength $0$, there are no prediction errors, and the results are identical for both cases.}
    \label{fig:per-step_errors}
\end{figure}

For our second set of experiments, we aim to provide evidence for the perturbation bounds presented in \cite{lin2022bounded}. However, this is not feasible because we do not know the exact values of $q_i(t)$ in the perturbation bounds. Instead of directly validating these bounds, we design experiments to assess the decay of the per-step error, which serves as a proxy for validating the perturbation bounds. If the perturbation bounds hold, we expect the per-step error to decay exponentially as a function of the prediction horizon $k$ when there is no prediction error, since the right-hand side of the perturbation bounds will approach zero. In the presence of prediction errors, the per-step error will still decay but will stabilize at some small value, as the right-hand sides of the perturbation bounds will converge to non-zero constants.

Figure~\ref{fig:per-step_errors} illustrates the average prediction errors for an episode of length $T=100$, under progressively increasing prediction horizons from 10 to 100, and varying levels of prediction errors with noise strength from 0.0 to 1.0. We observe that when there are no prediction errors, the per-step error decays exponentially as a function of the prediction horizon $k$. When only the disturbance is subject to prediction errors, the per-step error continues to decay over longer prediction horizons for noise strengths of 0.5 or lower, but it eventually stabilizes at a floor value. Similar results are observed when prediction errors affect all problem data. In the case where prediction errors are present across all problem data in the online MPC controller, the per-step error initially decays quickly before plateauing at very small values. However, when the noise strength exceeds 0.2, the optimization diverges, causing the per-step error to explode rather than remain small. This occurs because high prediction errors in the cost components related to control inputs can render the cost matrix $R_t(\xi_t)$ ill-conditioned, leading to the MPC controller selecting control inputs with increasingly large values.

From this experiment, we conclude that the per-step error behaves in a manner consistent with the perturbation bounds and the inequalities in Equation~\ref{lemma:pipeline-step2:conclusion}, thus providing proxy evidence supporting the perturbation bounds and the additional assumption proposed in \cite{lin2022bounded}.

\subsection{Exploring the Relationship Between Prediction Errors and Horizon for Dynamic Regret}

\begin{figure}[h!]
    \centering
    \includegraphics[width=0.9\textwidth]{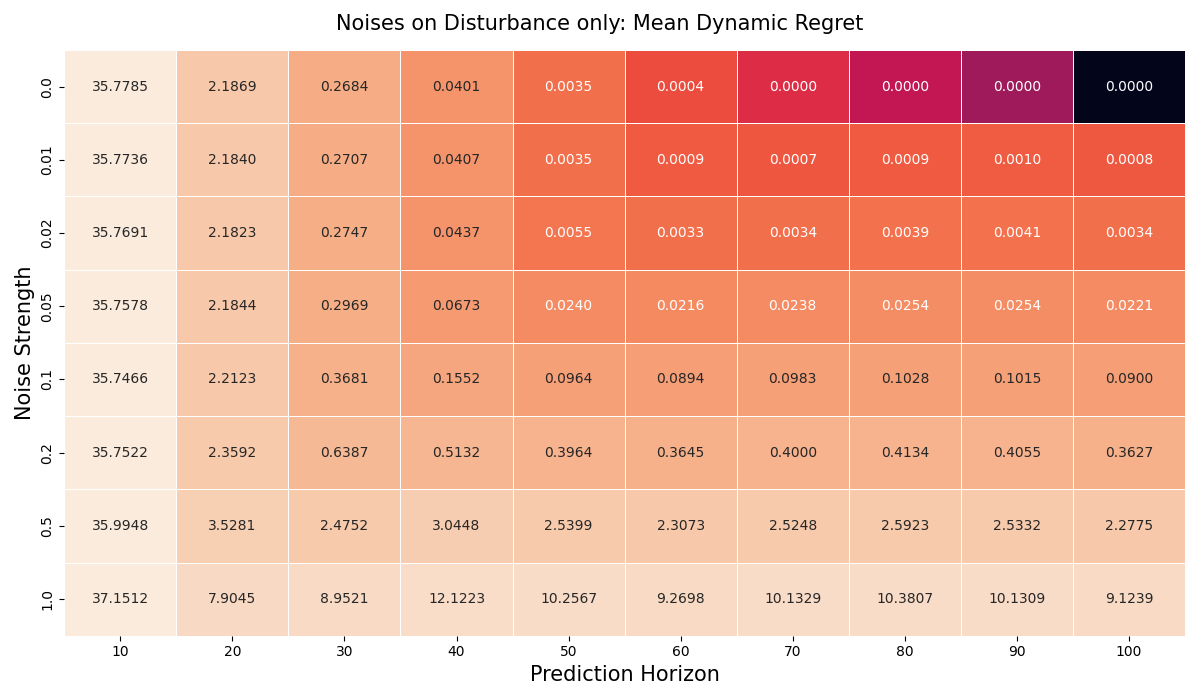} 
    \caption{Mean empirical dynamic regret over 5 runs for prediction errors on disturbance only.}
    \label{fig:regret_table_mean_disturbance} 
\end{figure}

\begin{figure}[h!]
    \centering
    \includegraphics[width=0.9\textwidth]{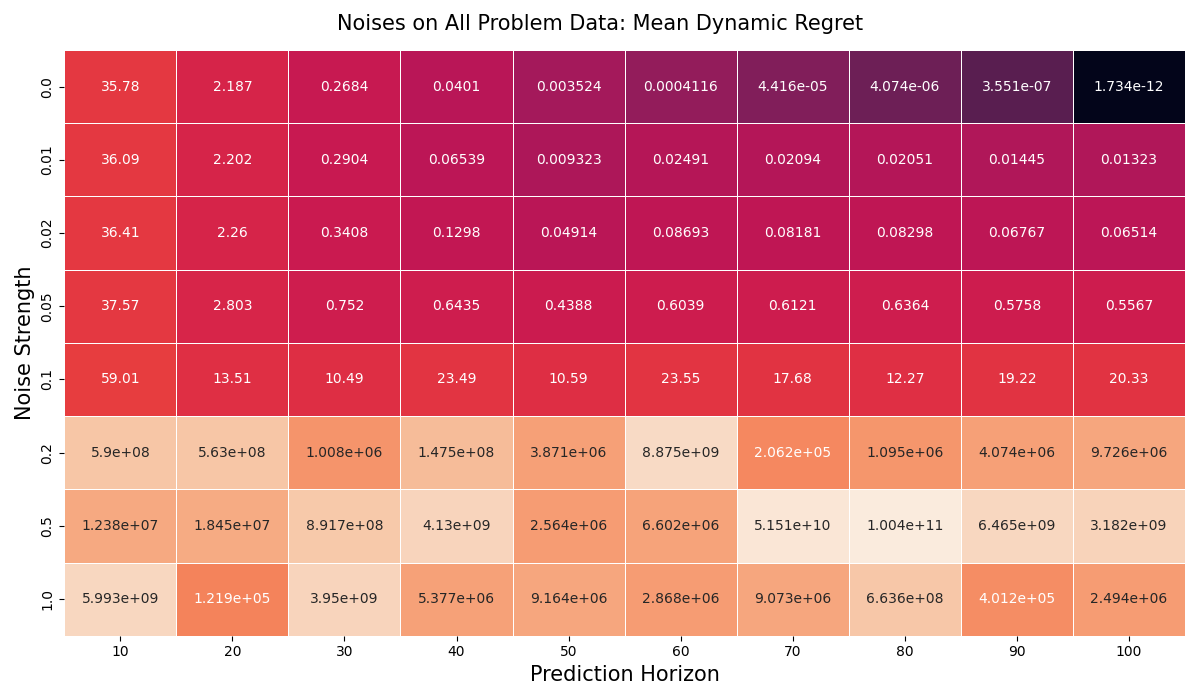} 
    \caption{Mean empirical dynamic regret over 5 runs for prediction errors on all problem data.}
    \label{fig:regret_table_mean_all} 
\end{figure}

\begin{figure}[h!]
    \centering
    \includegraphics[width=0.9\textwidth]{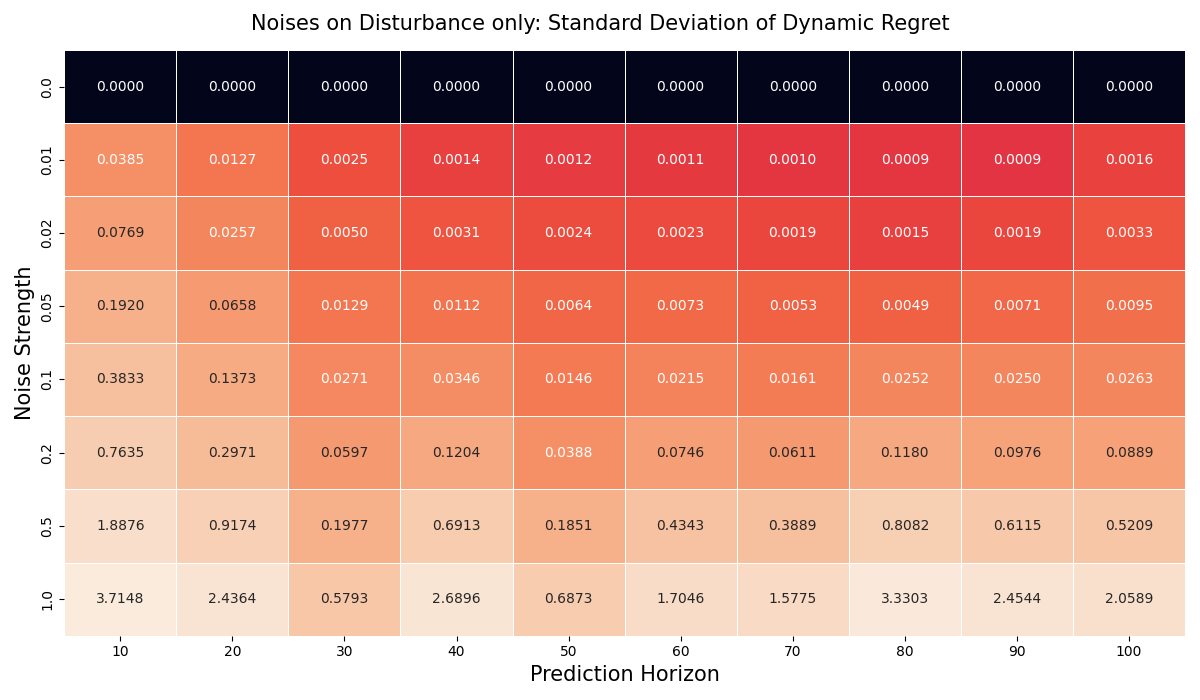} 
    \caption{Standard deviation of dynamic regret over 5 runs for prediction errors on disturbance only.}
    \label{fig:regret_table_std_disturbance} 
\end{figure}

\begin{figure}[h!]
    \centering
    \includegraphics[width=0.9\textwidth]{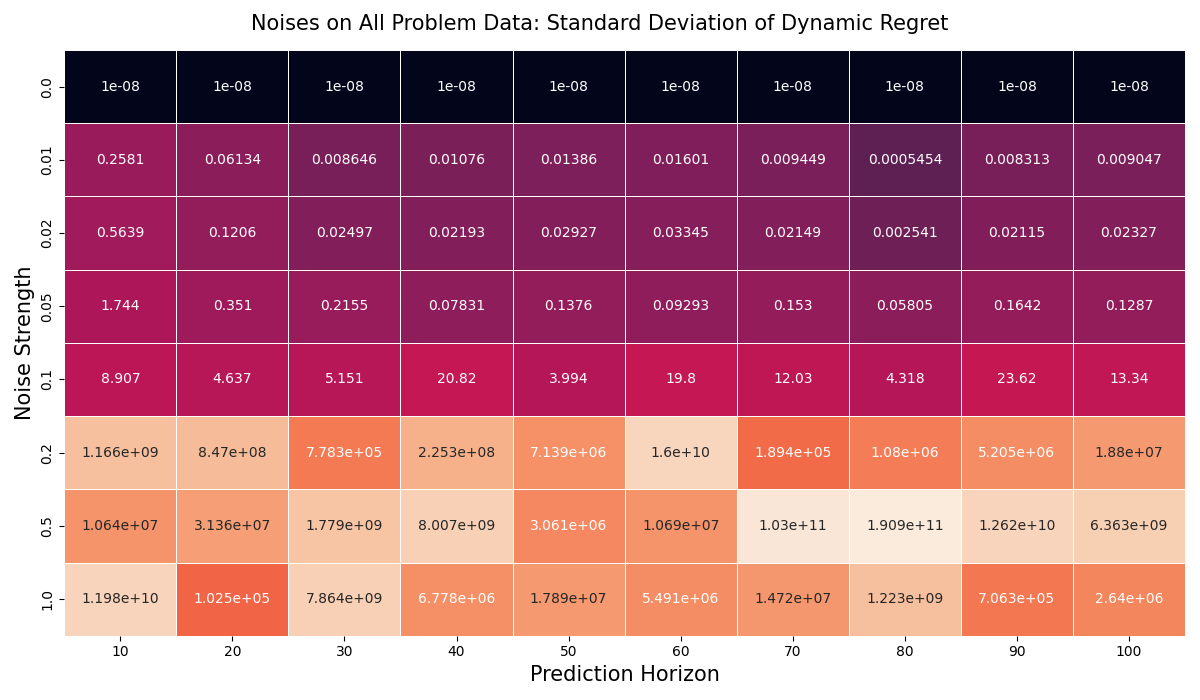} 
    \caption{Standard deviation of dynamic regret over 5 runs for prediction errors on all problem data.}
    \label{fig:regret_table_std_all} 
\end{figure}

In this section, we report the mean and standard deviation of the empirical dynamic regret for various levels of prediction errors and prediction horizons, based on five fixed random seeds at an episode length of $T=100$. The results are presented in the tables shown in Figures \ref{fig:regret_table_mean_disturbance}, \ref{fig:regret_table_mean_all}, \ref{fig:regret_table_std_disturbance}, and \ref{fig:regret_table_std_all}.

Figure \ref{fig:regret_table_mean_disturbance} illustrates that as prediction errors decrease and the prediction horizon increases, the mean empirical dynamic regret tends to decrease across the five random runs. In most cases, sublinear dynamic regret is achieved, except for the shortest prediction horizon of 10. This horizon is insufficient for effective MPC control, resulting in the highest dynamic regret. Additionally, for the highest prediction errors (noise strength of 1.0), the lowest dynamic regret occurs at a prediction horizon of 20. However, dynamic regret fluctuates and increases as the prediction horizon extends from 30 to 100. This behavior reflects a trade-off when using inaccurate predictions for online MPC: while a longer prediction horizon theoretically helps reduce dynamic regret, it also amplifies the inaccuracies in the predicted trajectory cost over multiple time steps. A similar effect is observed for a noise strength of 0.5. Thus, for each level of prediction error, there appears to be an optimal prediction horizon that minimizes dynamic regret, and this value may not always correspond to the longest horizon. This observation raises an interesting research question: What is the optimal prediction horizon for a specific level of prediction error in online MPC? This question is particularly relevant for resource-constrained systems where the prediction model cannot be trained to arbitrary precision.

Figure \ref{fig:regret_table_mean_all} shows similar results for the case when there are prediction errors on all problem data, though with higher dynamic regret overall. Additionally, we observe that when prediction errors exceed a certain threshold (noise strength of 0.2 or higher), the online MPC controller diverges. At these levels of prediction error, using MPC becomes ineffective.

Figures \ref{fig:regret_table_std_disturbance} and \ref{fig:regret_table_std_all} show the standard deviations of the runs. For the case where prediction errors are limited to the disturbance, the standard deviation is very low, indicating reliable results. In contrast, the standard deviation is higher when prediction errors affect all problem data, suggesting greater variability in the results.

\subsection{Neural Network-Based Prediction Model}

\begin{figure}[h!]
    \centering
    \includegraphics[width=0.7\textwidth]{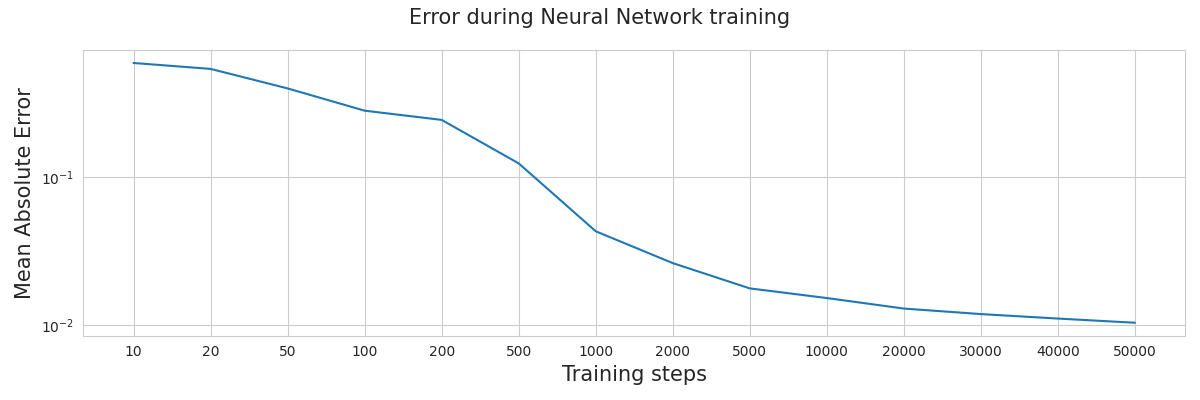} 
     \caption{Average prediction error over training steps for the neural network model.}
    \label{fig:nn_error} 
\end{figure}

For our final experiment, we evaluate the empirical dynamic regret using a commonly employed prediction model: a neural network with two fully connected hidden layers, each containing 256 units and using ReLU activation. The neural network takes the LQR system time step $\tilde{t}$ as input and outputs the problem data for the online MPC controller. We trained the neural network on all ground truth problem data for each time step, using 50,000 gradient steps. During the training process, the mean prediction errors of the neural network across all time steps are shown in Figure~\ref{fig:nn_error}. As seen, the neural network can learn to predict the problem data with high accuracy.

\begin{figure}[h!]
    \centering
    \includegraphics[width=0.9\textwidth]{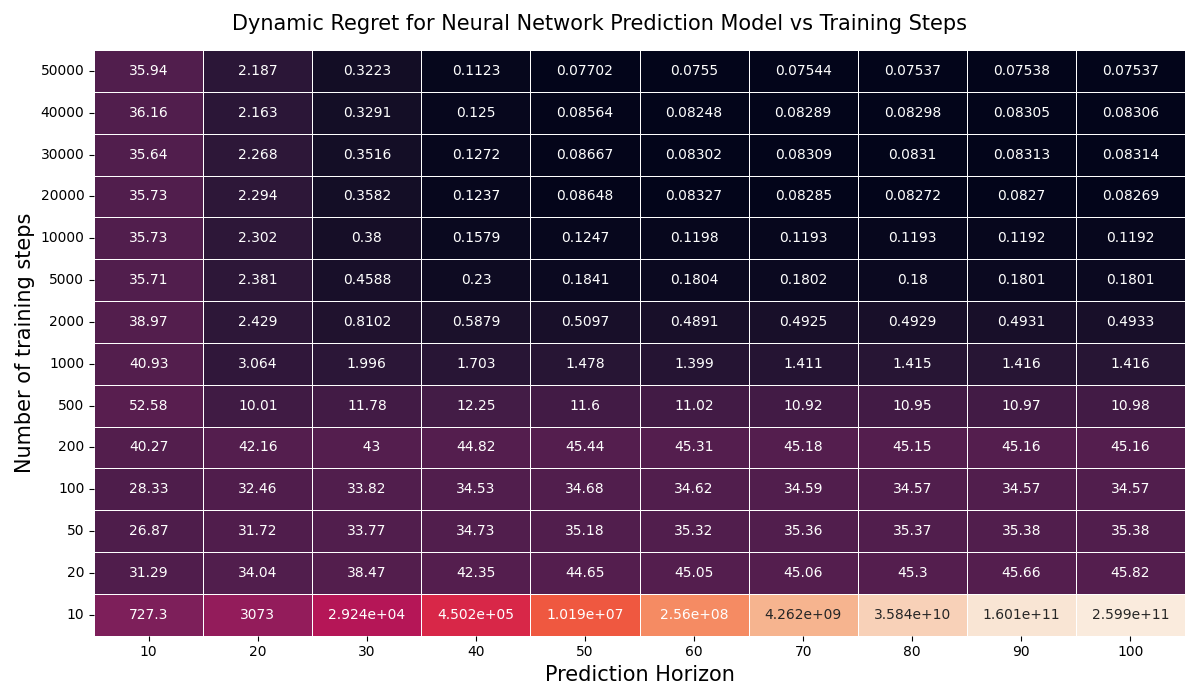} 
    \caption{Dynamic regret for the neural network prediction model across different training steps and prediction horizons.}
    \label{fig:regret_table_nn} 
\end{figure}

At various gradient steps ([10, 20, 50, 100, 200, 500, 1000, 2000, 5000, 10000, 20000, 30000, 40000, 50000]), the neural network's predictions are evaluated by using them in an online MPC controller, with prediction horizons $k$ ranging from 10 to 100. The corresponding empirical dynamic regret of the online MPC controller is presented in Figure~\ref{fig:regret_table_nn}. We observe that after only 500 gradient steps, the neural network can make sufficiently accurate predictions such that the empirical dynamic regret becomes sublinear. Furthermore, after just 20 gradient steps, the predictions improve enough to prevent the online MPC optimization from diverging or blowing up.

As with previous experiments, we note that a prediction horizon of 10 is too short to achieve sublinear dynamic regret, regardless of the prediction errors. Additionally, for gradient steps between 20 and 200, when prediction errors are still present but reasonable, increasing the prediction horizon can lead to poorer performance and higher dynamic regret. This reinforces the idea that, for resource-constrained systems, adaptively selecting the best prediction horizon based on prediction errors may be a promising area for future research.


\section{Code}
The code for this technical report is hosted on GitHub at this \href{https://github.com/minhnhat93/mpc-bounded-regret-evaluation/}{repository}. A brief explanation of the main files and directories:
\begin{itemize}
\itemsep -0.5em
    \item \textbf{"figures" directory}: contains all the figures that are used in this technical report.
    \item \textbf{ltv\_system.py}: code for the LQR system in section~\ref{section:LQR_formulation}. Also, contain. Also, contains code for controlling the prediction errors by adding uniform noises into MPC problem data.
    \item \textbf{mpc.py}: code for solving and collecting data for the online MPC controller.
    \item \textbf{nn\_prediction.py}: code for the neural network prediction model experiment.
    \item \textbf{run\_ltv\_ground\_truth\_parameters.py}: run and collect offline MPC controller data.
    \item \textbf{run\_ltv\_with\_param\_noise.py}: run and collect the online MPC controller data.
    \item \textbf{run\_ltv\_dynamic\_regret\_curve\_analysis.py}: run and collect empirical dynamic regret data.
    \item \textbf{run\_ltv\_with\_nn\_predictions.py}: run and collect data for the neural network experiment.
    \item \textbf{plot\_dynamic\_regret.py} and \textbf{plot\_per\_step\_error.py}: plot the figures in this technical report.
    \item \textbf{run\_cost\_correction.py}: correct the wrong cost for some of the data files.
\end{itemize}

\section{Conclusion and Future Directions}
This technical report provided empirical validation for the theoretical dynamic regret bound for online MPC in the work of \cite{lin2022bounded} for LTV systems. Specifically, we presented proxy evidence for the perturbation bounds, evidence for sublinear dynamic regret across multiple settings with varying prediction errors and horizons, and evidence for sublinear dynamic regret using a neural network as the prediction model. We also explored the interplay between prediction errors, prediction horizons, and dynamic regret. This investigation raises an interesting future research question for resource-constrained settings: how to adaptively select the best prediction horizon for online MPC under varying levels of prediction error. Another possible direction is empirical validation for nonlinear systems, which we were unable to address in this work.

\bibliographystyle{plainnat}
\bibliography{references}






\end{document}